\documentclass[12pt]{article}

\usepackage [latin1]{inputenc}
\usepackage[centertags]{amsmath}
\usepackage{amsmath}
\usepackage{amsfonts}
\usepackage{amssymb}
\usepackage{amsthm}
\usepackage{newlfont}
\usepackage{epsfig}
\usepackage{amscd}
\usepackage{pstricks}
\usepackage{epsfig}
\usepackage{graphicx}
\usepackage{color}
\usepackage{cite}

\newcommand{\beq}{\begin{equation}}
\newcommand{\eeq}{\end{equation}}
\newcommand{\ba}{\begin{array}}
\newcommand{\ea}{\end{array}}
\newcommand{\bea}{\begin{eqnarray}}
\newcommand{\eea}{\end{eqnarray}}

\usepackage[centertags]{amsmath}
\usepackage{amsfonts}
\usepackage{amssymb}
\usepackage{amsthm}
\usepackage{newlfont}
\usepackage{epsfig}
\usepackage{amscd}

\begin{document}

\begin{center}
{\large \sc \bf On the dispersionless Davey-Stewartson system: Hamiltonian vector fields Lax pair and relevant nonlinear Riemann-Hilbert problem for dDS-II system}

\vskip 20pt

{\large G. Yi  }

\vskip 20pt

{\it
School of Mathematics, Hefei University of Technology, Hefei 230601, China }

\bigskip

e-mail:  {\tt ge.yi@hfut.edu.cn}

\bigskip

{\today}

\end{center}

\begin{abstract}
In this paper, the semiclassical limit of Davey-Stewartson system is studied. It shows that the dispersionless limited integrable system of hydrodynamic type, which is defined as dDS (dispersionless Davey-Stewartson) system,  arises from the commutation condition of Lax pair of one-parameter vector fields.  The relevant nonlinear Riemann-Hilbert problem with reality constraint for the dDS-II system is also constructed. This kind of Riemann-Hilbert problem is meaningful for applying the formal inverse scattering transform method, recently developed by  Manakov and Santini, to study the dDS-II system .

\end{abstract}

\section{Introduction}
In 1974 Davey and Stewartson \cite{DavSte} used a multi-scale analysis to derive a coupled system of nonlinear partial differential equations describing the evolution of a three dimensional wave packet in water of a finite depth. The general DS(Davey-Stewartson) system (parametrized by $\varepsilon >0$) for a complex (wave-amplitude) field $q(x,y,t)$ and a real (mean-flow) field $\phi(x,y,t)$  is given by
\begin{subequations} \label{DavSte}
\begin{eqnarray}
\textbf{i}\varepsilon q_{t}+\frac{\varepsilon^{2}}{2}(q_{xx}+\sigma^{2} q_{yy})+\delta q \phi=0,
\end{eqnarray}
\begin{eqnarray}
\sigma^{2}\phi_{yy}-\phi_{xx}+(|q|^{2})_{xx}+\sigma^{2}(|q|^{2})_{yy}=0,
\end{eqnarray}
\end{subequations}
in which the variables $t,x,y \in \mathbb{R}$. We shall refer to (\ref{DavSte}) with $\sigma=1$ the DS-I (Davey-Stewartson-I) system and with $\sigma=\textbf{i}$ the DS-II (Davey-Stewartson-II) system. We also refer to (\ref{DavSte}) with $\delta=1$ the focusing case and  with $\delta=-1$ the defocusing case respectively.
The Davey-Stewartson system (\ref{DavSte}), as prototype example of classical integrable systems, has been extensively studied with many important results obtained \cite{AblCla}: $N$-line soliton solutions \cite{AnkFre,APPa,APPb,Naka1,Naka2,SatAbl}; localized exponentially decaying solitons \cite{BLMP}; an infinite dimensional symmetry group, in fact this involves an infinite dimensional Lie algebra with a Kac-Moody-Virasoro loop structure\cite{ChaWin,Omote}; similarity reductions to the second and fourth Painlev\'{e} equations \cite{Tajiri} ; a B\"{a}cklund transformation and Painlev\'{e} property\cite{GanLak}; an infinite number of commuting symmetries and conservation quantities, a recursion operator and bi-Hamiltonian structure\cite{FokSan,SanFok}.

As we know, the IST (inverse scattering transform) method is a powerful method to identify and solve
classes of integrable nonlinear PDEs and integrable dynamical systems. In
1967, Gardner, Greene, Kruskal and Miura pioneered this new method of
mathematical physics. They solved the Cauchy problem of the celebrated
KdV(Korteveg-de Vries) equation $u_{t}+uu_{x}+u_{xxx}=0$, a model equation for
 description of weakly nonlinear, weakly dispersive (1 + 1)-dimensional
waves, arising in many physical contexts, by making use of the ideas of direct and
inverse scattering for the stationary Schr\"{o}dinger operator $\hat{L}=-\partial_{x}^{2}+u(x,t)$ \cite{GGKM}.
In 1968, Lax generalized these ideas, showing in particular that integrable nonlinear PDEs arise from
commutation condition of linear partial differential operators (now called "Lax pairs") \cite{lax1}. In 1972, Zakharov
and Shabat showed that the IST method was not a particular method only for the KdV equation,
but it is also applicable to the NLS(nonlinear Schr\"{o}dinger) equation
$\textbf{i}q_{t}+q_{xx}+\delta |q|^{2}u=0$ \cite{zakharov1},
another important model equation in the description of the amplitude modulation of weakly nonlinear
and strongly dispersive waves in nature. Next, in 1973, Ablowitz, Kaup,
Newell and Segur developed a method to find a rather wide class of nonlinear evolution equations solvable
by these techniques \cite{AKNS1,AKNS2}, including the sine-Gordon equation
$u_{xt}=\sin u$. The classical IST method is the spectral method to solve the Cauchy problem for such PDEs, predicting that a localized disturbance evolves into
a number of  solitons (elastically interacting solitary waves arising from the balance of nonlinearity and dispersion) plus dispersive wave trains. The
soliton behavior has been observed in several physical contexts. These classical integrable systems are
often called "soliton PDEs". Physically
relevant integrable generalizations of the KdV and NLS equations to $(2+1)$ dimensions, respectively the KP(Kadomtsev-Petviashvili)
and DS(Davey-Stewartson) equations have also been constructed and solved
\cite{AblCla,Zakharov1984}. To the best of my knowledge, a further generalization to higher dimensions leads to non-integrable systems. Indeed,
except for few cases, it is impossible to construct nonlinear PDEs integrable by the classical IST
method in more than $2+1$ dimensions.

Besides the soliton PDEs, there is another important class of integrable PDEs: the "integrable PDEs of hydrodynamic type",
often called "dispersionless PDEs". In some cases, these PDEs are the dispersionless (semiclassical) limits of
integrable soliton equations. They often arise in various problems of \emph{Mathematical Physics} and are intensively studied in the recent literature.

Integrable nonlinear PDEs of hydrodynamic type are equivalent to the commutation
condition of vector fields Lax pairs.
For this reason, they could be in an arbitrary number of dimensions (as observed long ago by Zakharov and Shabat \cite{ZS0}).
In addition, since they do not contain dispersive or dissipative terms, an initial localized disturbance could evolve
towards a gradient catastrophe.
Therefore such integrable systems could give a good chance to obtain analytical results on the study of
wave breaking phenomena in multidimensions.

Manakov and Santini have introduced, at a formal level, a novel IST method for solving integrable
PDEs of hydrodynamic type, based
on the construction of a direct and inverse spectral problem for one-parameter families of vector fields
\cite{manakov1,manakov2}.
The particular nature of vector fields introduces important novelties with respect to the classical IST theory:

1) Since the space of the (zero energy) eigenfunctions of the vector fields
is a ring (not only the sum, but also the product of eigenfunctions is an
eigenfunction), the direct and inverse spectral theory is essentially nonlinear,
as opposed to those associated with the classical IST, in which the space of eigenfunctions is linear. In particular, the inverse problem can be
formulated as a nonlinear Riemann-Hilbert problem on a suitable contour of
the complex plane of the spectral parameter \cite{manakov1,manakov2}.

2) If the vector field Lax pair consists of Hamiltonian vector fields, the space of the eigenfunctions
is not only a ring, but also a Lie algebra, whose commutator is given by the Poisson bracket.

This novel IST method has been applied to solve the Cauchy problem for dKP (dispersionless Kadomtsev-Petviashvili) equation
\begin{eqnarray} \label{dispersionlessKP}
u_{xt}+u_{yy}+(uu_{x})_{x}=0,~~~~u=u(x,y,t) \in \mathbb{R},~~~~x,y,t \in \mathbb{R},
\end{eqnarray}
which is the dispersionless limit of the KP euqation. The dKP equation describes the evolution of small amplitude, nearly one-dimensional
waves with negligeable dispersion and dissipation. It appeared first in the description of unsteady motion in transonic flow \cite{timman1962} and in the nonlinear acoustics of confined beams \cite{zobolotskaya1969}.

The dKP equation arises from the commutation condition $[\hat{L}_{1},\hat{L}_{2}]=0$ of the pair of Hamiltonian vector fields
\begin{eqnarray}
\hat{L}_{1} &=& \partial_{y}+\lambda \partial_{x}-u_{x} \partial_{\lambda}=\partial_{y}+\{H_{2},\cdot\}_{(\lambda,x)}, \nonumber\\
\hat{L}_{2} &=& \partial_{t}+(\lambda^{2}+u) \partial_{x}-\lambda u_{x} \partial_{\lambda}=\partial_{t}+\{H_3,\cdot\}_{(\lambda,x)},
\end{eqnarray}
where $H_2,H_3$ are the following Hamiltonians
\begin{eqnarray}
H_2 &=&\frac{\lambda^2}{2}+u, \nonumber\\
H_{3} &=& \frac{\lambda^{3}}{3}+\lambda u-\partial^{-1}_{x}(u_{y}).
\end{eqnarray}
The dKP equation is the first nontrivial member (corresponding to $m=2,~n=3,~t_2=y,~t_3=t$) of the following hierarchy
of integrable PDEs
\begin{eqnarray}
{H_n,}_{t_m}-{H_m,}_{t_n}+\{H_m,H_n\}_{(\lambda,x)}=0,~~H_n\equiv \frac{1}{n}\left(f^n\right)_{\ge 0},
\end{eqnarray}
where
$f$ is the formal zero energy eigenfunction of the operator $\hat{L}_{1}$ (the solution of equation $\hat{L}_{1}f=0$) with a polar singularity in a neighborough of $\lambda=\infty$, with the expansion
\begin{eqnarray}
f=\lambda+\frac{u}{\lambda}-\frac{\partial^{-1}_{x}(u_{y})}{\lambda^{2}}+\sum_{j \geq 3,j \in \mathbb{Z}}\frac{q_{j}}{\lambda^{j}},~~|\lambda |\gg 1,
\end{eqnarray}
 where $\left( \right)_{\ge 0}$ stands for  the polynomial part of  Laurent expansion of $\lambda$ \cite{KG}. Here and hereafter in this paper, the \emph{Poisson bracket} is defined as follows
\begin{eqnarray}
\{A,B\}_{(s_{1},s_{2})}=A_{s_{1}} B_{s_{2}}-A_{s_{2}} B_{s_{1}}.
\end{eqnarray}

\iffalse
By using the IST method for vector fields, Manakov and Santini have constructed the formal solution of the Cauchy problems for the following novel system of PDEs
\begin{eqnarray} \label{MSsystem}
u_{xt}+u_{yy}+(uu_{x})_{x}+v_{x}u_{xy}-v_{y}u_{xx}&=&0, \nonumber\\
v_{xt}+v_{yy}+uv_{xx}+v_{x}v_{xy}-v_{y}v_{xx}&=&0
\end{eqnarray}
in \cite{manakov2}. The Cauchy problem for the v = 0 reduction of (\ref{MSsystem}), the dKP equation (\ref{dispersionlessKP}), was also
presented in \cite{manakov2}, while the Cauchy problem for the u = 0 reduction of (\ref{MSsystem}), an integrable
system introduced in \cite{Pavlov2003}, was given in \cite{Manakov2007}.
\fi

By using the IST method developed in \cite{manakov2}, Manakov and Santini \cite{manakov3} showed, in particular, that a
localized initial disturbance evolving according to dKP breaks at finite time and, if small, in the longtime regime, when the solution is described by the formula:
\begin{eqnarray} \label{ulongtime}
u=\frac{1}{\sqrt{t}}G(x+\frac{y^{2}}{4t}-2ut,\frac{y}{2t})+o(\frac{1}{\sqrt{t}}),~~t\gg 1,
\end{eqnarray}
where the spectral function $G$ is connected to the initial datum $u(x,y,0)$ via the direct spectral transform developed in \cite{manakov2}. According to formula (\ref{ulongtime}), a small and localised
initial datum evolves into a parabolic wave front described by equation  $x+\frac{y^{2}}{4t}=\tilde{x}$,
in the space-time region:
\begin{eqnarray}
x=\tilde{x}+v_{1}t,~~~y=v_{2}t,~~~\tilde{x}-2ut,v_{1},v_{2} =O(1),~~~v_{2} \neq 0,~~~t \gg 1,
\end{eqnarray}
and breaks in a point of the parabola. Indeed, since the argument of $G$ in (\ref{ulongtime}) depends
on $u$ itself through the combination $x+\frac{y^{2}}{4t}-2ut$, in full analogy with the case
of the general solution $v=v_0(x-vt)$ of the one - dimensional analogue of dKP, the celebrated Hopf equation $v_{t}+vv_{x}=0$, the breaking mechanism for dKP is very similar to that described
by the  Hopf equation, and
the details of such a $(2+1)$ dimensional wave breaking have been investigated analytically in a quite explicit manner. Therefore the dKP equation (\ref{dispersionlessKP}) could be viewed as a prototype
physical model equation in the description of wave breaking phenomena in $(2 + 1)$
dimensions, exactly as the Hopf equation is the
prototype model equation in the description of wave breaking phenomena in
$(1 + 1)$ dimensions.

The novel IST method has also been applied to solve the Cauchy problem for
the second heavenly equation of Pl\'{e}banski \cite{manakov1,manakov4}
\begin{eqnarray}\label{heavenly}
\theta_{xt} -\theta_{yz}+\theta_{xx}\theta_{yy}-\theta_{xy}^{2}=0,
\end{eqnarray}
an exact 4-dimensional reduction of the Einstein equations of General Relativity \cite{plebanski1}, and the
d2DT(dispersionless two-dimensional Toda) equation \cite{manakov5}
\begin{eqnarray}
\phi_{\zeta_{1}\zeta_{2}}=(e^{\phi_{t}})_{t},~~~~\phi=\phi(\zeta_{1},\zeta_{2},t),
\end{eqnarray}
whose elliptic and hyperbolic versions are relevant to integrable H-spaces (heavens) \cite{boyer1,gegenberg1} and integrable Einstein-Weyl geometries \cite{hitchin1,jones1,ward1}. The d2DT's string equations solutions are relevant in the ideal
Hele-Shaw problem \cite{mineev1,wigmann1,krichever1}.

 This novel IST method has recently been applied  to one distinguished class of equations, the so-called Dunajski hierarchy\cite{YiSan}. The Dunajski hierarchy is a basic example of dispersionless integrable PDEs, including the heavenly and the Manakov-Santini hierarchies as particular cases. The first flow of this hierarchy with the divergence free constraint is the well known Dunajski equation characterizing a general anti-self-dual conformal structure in neutral signature\cite{Dunaj}
\begin{subequations}\label{Dunaj}
\begin{eqnarray}
\theta_{xt} -\theta_{yz} + \theta_{xx} \theta_{yy} - \theta^{2}_{xy} = f,
\end{eqnarray}
\begin{eqnarray}
f_{xt} - f_{yz} + \theta_{yy} f_{xx} + \theta_{xx} f_{yy} -2 \theta_{xy} f_{xy}=0.
\end{eqnarray}
\end{subequations}

As the dKP equation is the integrable physically relevant generalization of the Hopf equation $u_t+uu_x=0$ in (2+1) dimensions, the
dDS(dispersionless Davey-Stewartson) system is the integrable physically relevant generalization of the dNLS(dispersionless nonlinear Schr\"{o}dinger) system of equations
\begin{subequations} \label{dNLS}
\begin{eqnarray}
u_{t}+(uv)_{x}=0,
\end{eqnarray}
\begin{eqnarray}
v_{t}+vv_{x}-\delta u_{x}=0, ~~~\delta=\pm 1,
\end{eqnarray}
\end{subequations}
The dNLS system is the dispersionless (semiclassical) limit $\varepsilon \rightarrow 0$ of the $\varepsilon$-dependent NLS equation
\begin{eqnarray} \label{eNLS}
\text{i}\varepsilon q_{t}+\frac{\varepsilon^{2}}{2} q_{xx}+\delta |q|^{2}q=0,  ~~\delta=\pm1,
\end{eqnarray}
where $u$ is the square modulus and $v$ is the wave number of $q$
\begin{eqnarray}
q=\sqrt{u}\exp(\textbf{i} \frac{\partial^{-1}_{x}(v)}{\varepsilon}),~u,v \in \mathbb{R},~u >0.
\end{eqnarray}
In the defocusing $\delta=-1$ case, the system is hyperbolic and describes an isentropic gas evolving towards a gradient catastrophe at finite time $t$ of the type described by the Hopf equation $u_{t}+uu_{x}=0$; in the focusing $\delta=1$ case,
the system is elliptic and evolves towards an elliptic umbilical singularity \cite{dubrovin2009}. Therefore
the dNLS system is clearly richer than the Hopf equation $u_{t}+uu_{x}=0$. It is reasonable to believe that the picture could be even richer in the $(2+1)$ dimensional case of the dDS system (\ref{dDS}). For this reason, we hope to apply the novel IST method to study the dDS system in a analytical way. Then we plan to study the wave breaking mechanism for the dDS system and to identify the type of singularities.

In fact, the defocusing DS-II system has been shown in numerical experiments to exhibit behavior in the semiclassical limit that qualitatively resembles that of its (1+1) dimensional reduction, the defocusing NLS equation \cite{KleRoi}. In 2017, Assainova, Klein,  Mclaughlin and Miller considered the direct spectral transforming for the defocusing DS-II system for smooth initial data in the semiclassical limit. They showed that the direct spectral transform involves a singularly-perturbed elliptic Dirac system in two dimensions. They introduced a WKB-type method for this problem, proved that it makes sense formally for sufficiently large values of the spectral parameter  by controlling the solution of an associated nonlinear eikonal problem. They also gave numerical evidence that the method is accurate for such parameter in the semiclassical limit \cite{AKMM}.

This paper is organized as follows: first, we construct the dDS system and its Hamilton-Jacobi Lax pair by taking the dispersionless (semiclassical) limit of the DS system (\ref{DavSte}) and of its Lax pair formulation; second, we derive its vector fields Lax pair formulation, the basic mathematical tool of the Manakov-Santini theory; third, we construct a nonlinear Riemann-Hilbert problem for dDS-II system with reality constraint.

By applying the IST method for vector fields, we plan to study
the Cauchy problem, to study how the dynamics give rise to a wave breaking, and to study analytically the nature of such wave breaking with the identification of the type of singularities. This paper is the essential starting point for the application of Manakov-Santini spectral theory.

\section{Semiclassical limit of Davey-Stewartson \\
system and Hamiltonian vector fields Lax pair  }
\subsection{Dispersionless Davey-Stewartson system and relevant Hamilton-Jacobi type equations}
 In this section, we consider the following $\varepsilon$-parametrized  DS(Davey-Stewartson) system
 \begin{subequations} \label{zDS}
 \begin{eqnarray}
 \textbf{i} \varepsilon q_{t}+\varepsilon^{2}(q_{zz}+q_{\hat{z}\hat{z}})+\delta q \phi=0,
 \end{eqnarray}
 \begin{eqnarray}
 \phi_{z\hat{z}}-\frac{1}{2}[(|q|^{2})_{zz}+(|q|^{2})_{\hat{z}\hat{z}}]=0,
 \end{eqnarray}
 \end{subequations}
 which is equivalent to the system (\ref{DavSte}) by introducing the independent variables
 \begin{eqnarray}
z=x+\sigma y, ~~~\hat{z}=x-\sigma y.
 \end{eqnarray}
 We refer to (\ref{zDS}) with $\delta=1$ the focusing case and  with $\delta=-1$ the defocusing case respectively.

The nonlinear DS system (\ref{zDS}) is equivalent to the compatibility condition of the following linear system
\begin{subequations}  \label{LaxDS}
\begin{eqnarray}
\varepsilon
\left(
\begin{array}{c}
  \psi_{z} \\
  \varphi_{\hat{z}}
\end{array}
\right)
=M
\left(
\begin{array}{c}
  \psi \\
  \varphi
\end{array}
\right),
\end{eqnarray}

\begin{eqnarray}
\varepsilon \left(
\begin{array}{c}
  \psi_{t} \\
  \varphi_{t}
\end{array}
\right)
=T
\left(
\begin{array}{c}
  \psi \\
  \varphi
\end{array}
\right),
\end{eqnarray}
\end{subequations}
where the two matrices $M$ and $T$ read as
\begin{eqnarray}
M=
\left(
  \begin{array}{cc}
    0 & -\frac{1}{2\sigma} q \\
    \frac{\sigma}{2} \delta \bar{q} & 0 \\
  \end{array}
\right),~~~~
T=\textbf{i}
 \left(
  \begin{array}{cc}
    \partial^{2}_{\hat{z}}+\frac{1}{2}\delta W & \frac{1}{2\sigma}(q\partial_{z}-q_{z}) \\
    \frac{\sigma}{2}(\bar{q} \partial_{\hat{z}}-\bar{q}_{\hat{z}}) & -\partial^{2}_{z}-\frac{1}{2}\delta V \\
  \end{array}
\right),
\end{eqnarray}
in which $\bar{q}$  is the complex conjugate of $q$ and
\begin{eqnarray}
W_{z}=(|q|^{2})_{\hat{z}},~~~V_{\hat{z}}=(|q|^{2})_{z},~~~\phi=\frac{1}{2}(W+V).
\end{eqnarray}

The well-known interpretation of the dispersionless (semiclassical) limit is that afforded by the quantum hydrodynamic system that one can derive from (\ref{zDS}) by following the ideas of Madelung \cite{Madelung}. Let us assume only that $|q| > 0$  and represent $q$ in the  "oscillatory wavepacket" or standard WKB form
\begin{eqnarray} \label{qS}
q=\sqrt{u}\exp(\textbf{i}\frac{S}{\varepsilon}),~~u>0,~~S \in \mathbb{R},
\end{eqnarray}
where $u(z,\hat{z},t) >0$ is a real amplitude and $S(z,\hat{z},t)$ is a real phase.
Inserting this form into (\ref{zDS}), dividing out the common phase factor and separating it into real and imaginary parts give the following system governing the three real-valued fields $u,S$ and $\phi$
\begin{subequations} \label{eDS}
\begin{eqnarray}
u_{t}+2(u S_{z})_{z}+2(u S_{\hat{z}})_{\hat{z}}=0,
\end{eqnarray}
\begin{eqnarray}
S_{t}+S^{2}_{z}+S^{2}_{\hat{z}}-\delta\phi=\varepsilon^{2}\frac{(\sqrt{u})_{zz}+(\sqrt{u})_{\hat{z}\hat{z}}}{\sqrt{u}},
\end{eqnarray}
\begin{eqnarray}
\phi_{z\hat{z}}-\frac{1}{2}(u_{zz}+u_{\hat{z}\hat{z}})=0.
\end{eqnarray}
\end{subequations}
In the dispersionless (semiclassical) limit $\varepsilon \to 0$,  the $\varepsilon$-dependent DS system (\ref{eDS}) reduces to the following dDS(dispersionless Davey-Stewartson) system
\begin{subequations} \label{dDS}
\begin{eqnarray}
u_{t}+2(u S_{z})_{z}+2(u S_{\hat{z}})_{\hat{z}}=0,
\end{eqnarray}
\begin{eqnarray}
S_{t}+S^{2}_{z}+S^{2}_{\hat{z}}-\delta\phi=0,
\end{eqnarray}
\begin{eqnarray}
\phi_{z\hat{z}}-\frac{1}{2}(u_{zz}+u_{\hat{z}\hat{z}})=0.
\end{eqnarray}
\end{subequations}

\bigskip
\bigskip
\textbf{Remark1.}
The uniform expression (\ref{dDS}) includes the dDS-I and dDS-II system. In fact, by considering the equations (\ref{dDS})  with the original
real-valued independent variables $(x,y,t)$, i.e., $u(z,\hat{z},t)=u(x+\sigma y,x-\sigma y,t)=,S(z,\hat{z},t)=S(x+\sigma y,x-\sigma y,t)$ (in this paper, we also use $u(x,y,t),S(x,y,t)$ to express the above two functions) ,
we obtain the following two different systems
\begin{subequations} \label{dDSreal}
\begin{eqnarray}
u_{t}+(u S_{x})_{x}+(u S_{y})_{y}=0,
\end{eqnarray}
\begin{eqnarray}
S_{t}+\frac{1}{2}(S^{2}_{x}+\sigma^{2} S^{2}_{y})-\delta \phi=0,
\end{eqnarray}
\begin{eqnarray}
\sigma^{2}\phi_{yy}-\phi_{xx}+u_{xx}+\sigma^{2} u_{yy}=0,
\end{eqnarray}
\end{subequations}
in which $\sigma=1$ corresponds the dDS-I system and $\sigma=\textbf{i}$ corresponds the dDS-II system.

~~~~~~~~~~~~~~~~~~~~~~~~~~~~~~~~~~~~~~~~~~~~~~~~~~~~~~~~~~~~~~~~~~~~~~~~~~~~~~~~~~~~~~~~~~~~~~~~~~~$\square$

To construct vector fields Lax pair for the dDS system (\ref{dDS}), by considering the requirement of phases equivalence in the system (\ref{LaxDS}), we follow Konopelchenko's method \cite{konopelchenko2007} and write the eigenfunctions $\psi$ and $\varphi$ of (\ref{LaxDS}) in the form
\begin{eqnarray}
\psi=\textbf{i}\psi_{0}\exp(\textbf{i}\frac{f}{\epsilon}),~~\varphi=\varphi_{0}\exp(\textbf{i}\frac{g}{\epsilon}),~~S=f-g.
\end{eqnarray}

By substituting these expressions into the linear system (\ref{LaxDS}), taking account of the terms of $O(\varepsilon^{0})$ and eliminating $\psi_{0}$, $\varphi_{0}$ from this system , one obtains the following system of three
nonlinear equations of Hamilton-Jacobi type
\begin{subequations} \label{fgLax}
\begin{eqnarray}
4f_{z}g_{\hat{z}}-\delta u=0,
\end{eqnarray}
\begin{eqnarray}
f_{t}+(f^{2}_{z}+f^{2}_{\hat{z}}-2f_{z}g_{z}-\frac{\delta}{2} W)=0,
\end{eqnarray}
\begin{eqnarray}
g_{t}-(g^{2}_{z}+g^{2}_{\hat{z}}-2f_{\hat{z}}g_{\hat{z}}-\frac{\delta}{2} V)=0,
\end{eqnarray}
\end{subequations}
where $S=f-g,~W_{z}=u_{\hat{z}},~V_{\hat{z}}=u_{z}$. Since the dDS system (\ref{dDS}) arises from the compatibility
condition
of equations (\ref{fgLax}), the equations (\ref{fgLax}) should be interpreted as the nonlinear Lax formulation of Hamilton-Jacobi type of dDS system.

\subsection{Hamiltonian vector fields Lax pair for \\
dispersionless Davey-Stewartson system}
In order to apply the Manakov-Santini method, it is important to construct an alternative vector field formulation of such a Lax pair. In this subsection, we construct this kind of Hamiltonian vector fields Lax pair.

As we know, some basic facts exist in the classical Hamiltonian mechanics in the real framework. Suppose
\begin{eqnarray}
\tilde{L}=\partial_{t}-\{\tilde{H},\cdot \}_{(p,x)},~~~t,x,p \in \mathbb{R},
\end{eqnarray}
is a Hamiltonian vector field, with the Hamiltonian
\begin{eqnarray}
\tilde{H}=\tilde{H}(t,x,p).
\end{eqnarray}
Any eigenfunction $\tilde{\Psi}(t,x,p)$ of $\tilde{L}$, i.e.,
\begin{eqnarray}
\tilde{L}\tilde{\Psi}=0,~~~~ t,x,p \in \mathbb{R},~~\tilde{\Psi} \in \mathbb{R},
\end{eqnarray}
is exactly a conservation law for the associated dynamical system
\begin{eqnarray}
\frac{dx}{dt}=\tilde{H}_{p}(t,x,p),~~~\frac{dp}{dt}=-\tilde{H}_{x}(t,x,p).
\end{eqnarray}

Hamiltonian systems can also be studied using the Hamilton-Jacobi equation
\begin{eqnarray}
\frac{\partial \tilde{K}} {\partial t}=\tilde{H}(t,x,\frac{\partial \tilde{K}}{\partial x}).
\end{eqnarray}

In classical Hamiltonian mechanics, above expressions give three equivalent formulations of the problem. These facts, especially the connection between vector fields and Hamiltonian-Jacobi equations, give us a way to construct vector fields Lax pair based on some Hamiltonian-Jacobi equations, even if the variables are complex.

Let's consider the Hamiltonian vector field
\begin{eqnarray}
L=\partial_{s}-\{H,\cdot\}_{(\lambda,z)},~~~~s,z,\lambda \in \mathbb{C},
\end{eqnarray}
with the Hamiltonian
\begin{eqnarray}
H=H(s,z,\lambda).
\end{eqnarray}

For any eigenfunction $\Psi(s,z,\lambda)$ of $L$, i.e., $L\Psi=0$, by considering the level sets
\begin{eqnarray}  \label{psik}
 \Psi(s,z,\lambda)=k
\end{eqnarray}
and solving this relation at the regular points, i.e., $\Psi_{\lambda}(s,z,\lambda) \neq 0$
\begin{eqnarray}  \label{Lambda}
\lambda=\Lambda(s,z,k),
\end{eqnarray}
we could define a complex value function $\Lambda(s,z,k)$. The following lemma shows the connection between vector field (complex variables) and Hamiltonian-Jacobi equation is similar as Hamilton mechanics.

\textbf{Lemma1.} The complex-valued function  $\Lambda(s,z,k)$ , connected with the Hamiltonian vector field $L$ by the above formula (\ref{psik})(\ref{Lambda}), satisfies the equation
\begin{eqnarray}
\Lambda_{s}=\frac{\partial}{\partial z} [H\left(s,z,\Lambda(s,z,k)\right)].
\end{eqnarray}
If we define $\Lambda(s,z,k) \equiv K_{z}(s,z)$, then $K(s,z)$ satisfies the Hamilton-Jacobi equation
\begin{eqnarray}
K_{s}=H(s,z,K_{z}).
\end{eqnarray}

\textbf{Proof.} The condition $\Psi(s,z,\lambda)$ is an eigenfunction of vector field $L$, which means
\begin{eqnarray} \label{lemma1}
L\Psi=\Psi_{s}-H_{\lambda} \Psi_{z}+H_{z}\Psi_{\lambda}=0.
\end{eqnarray}
The relations (\ref{psik}) and (\ref{Lambda}) read as
\begin{eqnarray} 
\Psi(s,z,\Lambda(s,z,k))=k,
\end{eqnarray}
which leads to
\begin{eqnarray} \label{lemma2}
0&=& \frac{\partial k}{\partial s}=\Psi_{s}+\Psi_{\lambda} \Lambda_{s}, \nonumber\\
0&=& \frac{\partial k}{\partial z}=\Psi_{z}+\Psi_{\lambda} \Lambda_{z}.
\end{eqnarray}
By considering both (\ref{lemma1}) and (\ref{lemma2}), one obtains the following relation
\begin{eqnarray}
0&=&\Psi_{s}+\Psi_{\lambda} \Lambda_{s}-H_{\lambda}(\Psi_{z}+\Psi_{\lambda} \Lambda_{z}) \nonumber\\
&=& (\Lambda_{s}-H_{\lambda} \Lambda_{z}-H_{z})\Psi_{\lambda}.
\end{eqnarray}
Taking into account that $\Psi_{\lambda} \neq 0$, one obtains
\begin{eqnarray}
\Lambda_{s}&=&H_{\lambda} \Lambda_{z}+H_{z} \nonumber\\
&=&\frac{\partial }{\partial z}[H\left(s,z,\Lambda(s,z,k)\right)].
\end{eqnarray}

~~~~~~~~~~~~~~~~~~~~~~~~~~~~~~~~~~~~~~~~~~~~~~~~~~~~~~~~~~~~~~~~~~~~~~~~~~~~~~~~~~~~~~~~~~~~~~~~~~~$\square$

\bigskip
\bigskip
\bigskip

By taking advantage of Lemma 1, one obtains the following main result in this section.

\textbf{Proposition 1.} The dDS system (\ref{dDS}) arises from the commutation condition
\begin{eqnarray} \label{commu}
[L_{1},L_{2}]=0
\end{eqnarray}
of the following one-parameter Hamiltonian vector fields Lax pair
\begin{subequations} \label{vecLax}
\begin{eqnarray}
L_{1}=\partial_{\hat{z}}-\{H_{1},\cdot\}_{(\lambda,z)},
\end{eqnarray}
\begin{eqnarray}
L_{2}=\partial_{t}-\{H_{2},\cdot\}_{(\lambda,z)},
\end{eqnarray}
\end{subequations}
with the Hamiltonian functions
\begin{subequations} \label{Hamilt}
\begin{eqnarray}
H_{1}(\lambda)=S_{\hat{z}}+\frac{\delta}{4} \frac{u}{\lambda},
\end{eqnarray}
\begin{eqnarray}
H_{2}(\lambda)=\lambda^{2}-2 S_{z} \lambda+(\frac{\delta}{2} W-S^{2}_{\hat{z}})-\frac{\delta}{2} \frac{u S_{\hat{z}}}{\lambda}-\frac{1}{16} \frac{u^{2}}{\lambda^{2}},
\end{eqnarray}
\end{subequations}
in which the parameter $\lambda \in \mathbb{C}/\{0\}$.

\textbf{Proof.} The commutation condition $[L_{1},L_{2}]=0$ is equivalent to the condition that the two vector fields $L_{1}$ and $L_{2}$ share the same eigenfunctions $\Psi(t,\hat{z},z,\lambda)$, i.e., $L_{1}\Psi=0 \Leftrightarrow  L_{2}\Psi=0$. By using the conclusion of Lemma 1, one obtains the following relations
\begin{subequations}
\begin{eqnarray}
\Lambda_{\hat{z}}= \frac{\partial}{\partial z}[H_{1}(\Lambda)]= \frac{\partial}{\partial z}[S_{\hat{z}}+\frac{\delta}{4} \frac{u}{\Lambda}]
\end{eqnarray}
\begin{eqnarray}
\Lambda_{t}= \frac{\partial}{\partial z}[H_{2}(\Lambda)]= \frac{\partial}{\partial z}\left[\Lambda^{2}-2 S_{z} \Lambda+(\frac{\delta}{2} W-S^{2}_{\hat{z}})-\frac{\delta}{2} \frac{u S_{\hat{z}}}{\Lambda}-\frac{1}{16} \frac{u^{2}}{\Lambda^{2}}\right],
\end{eqnarray}
\end{subequations}
in which $\Psi\left(t,\hat{z},z,\Lambda(t,\hat{z},z,k)\right)=k$.

By noting $f_{z}(t,\hat{z},z)=\Lambda(t,\hat{z},z,k)=\lambda$, one obtains the following equations
\begin{subequations}
\begin{eqnarray}
f_{\hat{z}}= H_{1}(f_{z})= S_{\hat{z}}+\frac{\delta}{4} \frac{u}{f_{z}},
\end{eqnarray}
\begin{eqnarray}
f_{t}= H_{2}(f_{z})= f_{z}^{2}-2 S_{z} f_{z}+(\frac{\delta}{2} W-S^{2}_{\hat{z}})-\frac{\delta}{2} \frac{u S_{\hat{z}}}{f_{z}}-\frac{1}{16} \frac{u^{2}}{f_{z}^{2}},
\end{eqnarray}
\end{subequations}
which are equivalent to the Hamilton-Jacobi equations (\ref{fgLax}), leading to  the dDS system (\ref{dDS}).
~~~~~~~~~~~~~~~~~~~~~~~~~~~~~~~~~~~~~~~~~~~~~~~~~~~~~~~~~~~~~~~~~~~~~~~~~~~$\square$

\bigskip
\bigskip
\textbf{Remark 2.} Since $L_{1}$ and $L_{2}$ are Hamiltonian vector fields, the commutation condition (\ref{commu}) is equivalent to
\begin{eqnarray} \label{zero}
H_{1,t}-H_{2,\hat{z}}+\{H_{1},H_{2}\}_{(\lambda,z)}=0.
\end{eqnarray}
Then the dDS system(\ref{dDS}) is equivalent to this Zakharov-Shabat equation (\ref{zero}) for $H_{1}$ and $H_{2}$.

~~~~~~~~~~~~~~~~~~~~~~~~~~~~~~~~~~~~~~~~~~~~~~~~~~~~~~~~~~~~~~~~~~~~~~~~~~~~~~~~~~~~~~~~~~~~~~~~~~~$\square$

\textbf{Remark 3.} An integrable PDE system is usually associated with a hierarchy of PDEs defining infinitely many
symmetries. A new hierarchy related to dDS system, which is so called the dDS hierarchy, can also be defined \cite{yi2019}.

~~~~~~~~~~~~~~~~~~~~~~~~~~~~~~~~~~~~~~~~~~~~~~~~~~~~~~~~~~~~~~~~~~~~~~~~~~~~~~~~~~~~~~~~~~~~~~~~~~~$\square$

\textbf{Proposition 2.} Consider the following nonlinear vector Riemann-Hilbert problem
\begin{eqnarray} \label{NRH1}
\xi^{+}_{j}(\lambda)=\xi^{-}_{j}(\lambda)+R_{j}(\xi^{-}_{1}(\lambda)+\mu_{1}(\lambda),\xi^{-}_{2}(\lambda)+\mu_{2}(\lambda)),
\lambda \in \Gamma, j=1,2
\end{eqnarray}
on an arbitrary closed contour $\Gamma$ of the complex $\lambda$- plane including the original point, where $\vec{R}(\vec{s})=(R_{1}(s_{1},s_{2}),R_{1}(s_{1},s_{2}))^{T}$ are given differentiable spectral data satisfying the constraint
\begin{eqnarray} \label{constraint1}
\{\mathcal{R}_{1}(s_{1},s_{2}),\mathcal{R}_{2}(s_{1},s_{2})\}_{(s_{1},s_{2})}=1,\nonumber\\
\mathcal{R}_{j}(s_{1},s_{2}) \equiv s_{j}+R_{j}(s_{1},s_{2}),~~~j=1,2
\end{eqnarray}
and $\mu_{j}(j=1,2)$ are the explicit functions
\begin{eqnarray}
\vec{\mu}=\left(\begin{array}{c}
            \mu_{1} \\
            \mu_{2}
          \end{array} \right)
          =\left(\begin{array}{c}
            \textbf{i}(\lambda-S_{z}+\frac{\delta}{4} \frac{u}{\lambda}) \\
           t(\lambda-S_{z}-\frac{\delta}{4} \frac{u}{\lambda})+\frac{z}{2}
          \end{array} \right).
\end{eqnarray}
 The vectors $\vec{\xi}^{+}=(\xi^{+}_{1},\xi^{+}_{2})^{T},\vec{\xi}^{-}=(\xi^{-}_{1},\xi^{-}_{2})^{T}$ are unknown vector solutions of the Riemann-Hilbert problem (\ref{NRH1}), analytic respectively inside and outside the contour $\Gamma$ such that $\vec{\xi}^{-} \to \vec{0}$ as $\lambda \to \infty$.

If the above nonlinear Riemann-Hilbert problem (\ref{NRH1}) and its linearized form  are uniquely solvable, and if the solutions of (\ref{NRH1}) satisfy the following closure conditions
\begin{subequations} \label{closure1}
\begin{eqnarray}
u=\lim_{\lambda \to \infty} 2\delta \lambda\left[\textbf{i}\xi^{-}_{1}(\lambda)+\frac{1}{t} \xi^{-}_{2}(\lambda)\right],
\end{eqnarray}
\begin{eqnarray}
S_{z}=\frac{z-\hat{z}}{4t}-\frac{\textbf{i}}{2} \xi^{+}_{1}(0)+\frac{1}{2t} \xi^{+}_{2}(0),
\end{eqnarray}
\end{subequations}
then $\vec{\pi}^{\pm}=\vec{\xi}^{\pm}+\vec{\mu}$ are common eigenfunctions of the vector fields (\ref{vecLax}):$L_{j} \vec{\pi}^{\pm}=\vec{0}~(j=1,2)$ satisfying the relations
\begin{eqnarray}
\{\pi^{\pm}_{1},\pi^{\pm}_{2}\}_{(\lambda,z)}=\frac{\textbf{i}}{2}
\end{eqnarray}
and the potentials $u,S_{z}$ reconstructed through (\ref{closure1}) solve the dDS system (\ref{dDS}).

\textbf{Proof.} The Riemann-Hilbert problem (\ref{NRH1}) could be formulated directly in terms of the eigenfunctions $\vec{\pi}^{\pm}=\vec{\xi}^{\pm}+\vec{\mu}$ as follows
\begin{eqnarray}  \label{NRHeigen1}
\pi^{+}_{j}(\lambda)&=&\mathcal{R}_{j}\left(\pi^{-}_{1}(\lambda),\pi^{-}_{2}(\lambda)\right) \nonumber\\
&=&\pi^{-}_{j}(\lambda)+R_{j}\left(\pi^{-}_{1}(\lambda),\pi^{-}_{2}(\lambda)\right),\lambda \in \Gamma, j=1,2
\end{eqnarray}
with the normalization
\begin{eqnarray} \label{normalization1}
\vec{\pi}^{-}(\lambda)=\vec{\mu}(\lambda)+O(\lambda^{-1}),~~~|\lambda| >>1.
\end{eqnarray}
By applying the operators  $L_{j}~ (j=1,2)$  defined in (\ref{vecLax}) to the nonlinear Riemann-Hilbert problem (\ref{NRHeigen1}), one obtains the linearized Riemann-Hilbert problem
\begin{eqnarray}
L_{j}\vec{\pi}^{+}(\lambda)=J L_{j}\vec{\pi}^{-}(\lambda),~~\lambda \in \Gamma,
\end{eqnarray}
where $J$ is the Jacobian matrix defined as $J_{ij}=\partial \mathcal{R}_{i}/\partial s_{j}~(i,j=1,2).$ Based on the normalization (\ref{normalization1}), we have $L_{j}\vec{\pi}^{-} \to \vec{0}$ when $\lambda \to \infty$. It follows that with the uniquely solvable assumption, $\vec{\pi}^{\pm}$ are common eigenfunctions of the vector fields, i.e. ,
$L_{j} \vec{\pi}^{\pm}=\vec{0}~(j=1,2).$    Consequently, the potentials $u,S$  solve the dDS system (\ref{dDS}).

Then the eigenfunctions exhibit the following asymptotics
\begin{subequations}
\begin{eqnarray}
\pi^{-}_{1}=\textbf{i}(\lambda-S_{z}-\frac{\delta}{4} \frac{V}{u})+O(\lambda^{-2}),~~~~|\lambda| >>1,
\end{eqnarray}
\begin{eqnarray}
\pi^{+}_{1}=-\textbf{i}(-\frac{\delta}{4} \frac{u}{\lambda}-S_{\hat{z}}+\frac{W}{u})\lambda+O(\lambda^{2}),~~~~|\lambda| <<1,
\end{eqnarray}
\begin{eqnarray}
\pi^{-}_{2}=\lambda t+(\frac{z}{2}-t S_{z})-\frac{\delta}{4} \frac{t V}{\lambda}+O(\lambda^{-2}),~~~~|\lambda| >>1,
\end{eqnarray}
\begin{eqnarray}
\pi^{-}_{2}=-\frac{\delta}{4}\frac{tu}{\lambda}+(\frac{\hat{z}}{2}-t S_{\hat{z}})+ \frac{tW}{u} \lambda+O(\lambda^{2}),~~~~|\lambda| <<1,
\end{eqnarray}
\end{subequations}
with $W_{z}=u_{\hat{z}},~V_{\hat{z}}=u_{z}$, implying the closure conditions (\ref{closure1}) which can be viewed as a nonlinear system of two algebraic equations for $u$ and $S_{z}$.

Based on the relation $\{\pi^{+}_{1},\pi^{+}_{2}\}_{(\lambda,z)}=\{\mathcal{R}_{1},\mathcal{R}_{2}\}_{(\pi^{-}_{1},\pi^{-}_{2})} \{\pi^{-}_{1},\pi^{-}_{2}\}_{(\lambda,z)},\lambda \in \Gamma$ from (\ref{NRHeigen1}), the constraint (\ref{constraint1}) implies that
$\{\pi^{+}_{1},\pi^{+}_{2}\}_{(\lambda,z)}=\{\pi^{-}_{1},\pi^{-}_{2}\}_{(\lambda,z)},\lambda \in \Gamma$, i.e. the Poisson brackets $\{\pi^{+}_{1},\pi^{+}_{2}\}_{(\lambda,z)},\{\pi^{-}_{1},\pi^{-}_{2}\}_{(\lambda,z)}$ are analytic in the whole complex $\lambda$-plane. As $\{\pi^{-}_{1},\pi^{-}_{2}\}_{(\lambda,z)} \to \frac{\textbf{i}}{2}$ when $\lambda \to \infty$, it follows that $\{\pi^{+}_{1},\pi^{+}_{2}\}_{(\lambda,z)}=\{\pi^{-}_{1},\pi^{-}_{2}\}_{(\lambda,z)}=\frac{\textbf{i}}{2}$.

~~~~~~~~~~~~~~~~~~~~~~~~~~~~~~~~~~~~~~~~~~~~~~~~~~~~~~~~~~~~~~~~~~~~~~~~~~~~~~~~~~~~~~~~~~~~~~~~~~~$\square$

\textbf{Remark 4.} This proposition provides us with a general way to construct the complex-valued solutions to the dDS system (\ref{dDS}) through a nonlinear Riemann-Hilbert problem. But in fact the real-valued restriction  is required by the original physics meaning (\ref{qS}). In order to guarantee this, we need to give the above nonlinear Riemann-Hilbert problem some constraints. We will show this result for dDS-II system in the following section.

~~~~~~~~~~~~~~~~~~~~~~~~~~~~~~~~~~~~~~~~~~~~~~~~~~~~~~~~~~~~~~~~~~~~~~~~~~~~~~~~~~~~~~~~~~~~~~~~~~~$\square$

\section{The relevant nonlinear Riemann-Hilbert\\
 problem for dDS-II system}

 In order to apply the Manokov-Santini novel IST method to study the dDS system, similarly as the previously mentioned hydrodynamical systems, such as the dKP equation, the Pavlov equation, the second heavenly equation of Pl\'ebanski and the d2DT (dispersionless 2D Toda) equation,  it is important to construct relevant nonlinear Riemann-Hilbert problems in the complex $\lambda$ plane with some constraints. Particularly, the reality constraint is important since it gives us the real-valued solutions with clear physics significance.

 In the d2DT case, also in this dDS framework, $0$ and $\infty$ are the singular points in the complex $\lambda$ plane. In the d2DT case, the generators of the two Hamiltonians are two independent eigenfunctions respectively, one with polar singularity around $0$ and the other around $\infty$. But now in this dDS case, both eigenfunctions appear in the construction of the Hamiltonian. Therefore the nonlinear Riemann-Hilbert problem associated with the dDS system is more complicated than that associated with the dKP and the d2DT.
 When $\sigma=\textbf{i}$, independent variables $z,\hat{z}$ read as conjugate complex variables, this symmetry helps us to construct a symmetric nonlinear Riemann-Hilbert problem in the complex plane. For this reason, we study dDS-II system (with $\sigma=\textbf{i}$) in this section, and note $\bar{z} := \hat{z}$ means the conjugate of $z$. Construction of the relevant nonlinear Riemann-Hilbert problem for dDS-II system with reality constraint will be demonstrated in this section.

Proposition 1 shows that the dDS system (\ref{dDS}) arises from the commutation condition $[L_{1},L_{2}]=0$. As we already know, this commutation condition is equivalent to the condition that the two vector fields $L_{1}$ and $L_{2}$ share the same eigenfunctions $\Psi(t,\bar{z},z,\lambda)$, i.e., $L_{1}\Psi=0 \Leftrightarrow  L_{2}\Psi=0$. Now we introduce a new parameter $p$ (here and hereafter $\sqrt{u}$ means the principal value)
\begin{eqnarray}
p=\frac{2}{\sqrt{u}} \lambda,
\end{eqnarray}
and define
\begin{eqnarray}
\Psi(t,\bar{z},z,\lambda)=\Psi(t,\bar{z},z,\frac{\sqrt{u}}{2} p)=\Phi(t,\bar{z},z,p).
\end{eqnarray}

Then the description of common eigenfunctions for $L_{1},L_{2}$, i.e., $L_{1}\Psi(\lambda)=0$ and $L_{2}\Psi(\lambda)=0$, reads as follows
\begin{subequations}
\begin{eqnarray}
\frac{2}{\sqrt{u}} \mathcal{L}_{1}[p] \Phi(p) =0,
\end{eqnarray}
\begin{eqnarray}
\frac{2}{\sqrt{u}} \mathcal{L}_{2}[p] \Phi(p) =0,
\end{eqnarray}
\end{subequations}
where
\begin{subequations} \label{mathcalL}
\begin{eqnarray}
\mathcal{L}_{1}[p]=\{\frac{\sqrt{u}}{2} p,\cdot\}_{(p,\bar{z})}-\{H_{1}(\frac{\sqrt{u}}{2} p),\cdot\}_{(p,z)}~ ,
\end{eqnarray}
\begin{eqnarray}
\mathcal{L}_{2}[p]=\{\frac{\sqrt{u}}{2} p,\cdot\}_{(p,t)}-\{H_{2}(\frac{\sqrt{u}}{2} p),\cdot\}_{(p,z)} ~,
\end{eqnarray}
\end{subequations}
in which the functions $H_{1}(\lambda)$ and $H_{2}(\lambda)$ are defined in (\ref{Hamilt}).

\textbf{Remark 5.} The commutation condition $[L_{1},L_{2}]=0$, which is equivalent to the dDS-II system, reads as
$\left[\frac{2}{\sqrt{u}} \mathcal{L}_{1}[p],\frac{2}{\sqrt{u}} \mathcal{L}_{2}[p]\right]=0$ (instead of \\ $\left[\mathcal{L}_{1}[p],\mathcal{L}_{2}[p]\right]=0$ ).   And this commutation condition is also equivalent to the condition that the two vector fields $\mathcal{L}_{1}[p]$ and $\mathcal{L}_{2}[p]$ share the same eigenfunctions $\Phi(p)$, i.e., $\mathcal{L}_{1}[p]\Phi(p)=0 \Leftrightarrow  \mathcal{L}_{2}[p]\Phi(p)=0$.

~~~~~~~~~~~~~~~~~~~~~~~~~~~~~~~~~~~~~~~~~~~~~~~~~~~~~~~~~~~~~~~~~~~~~~~~~~~~~~~~~~~~~~~~~~~~~~~~~~~$\square$

By taking advantage of  the new parameter $p$, one obtains the following Proposition 3 for constructing the real solutions $u,S$ for dDS-II system.

\textbf{Proposition 3.} Consider the following nonlinear vector Riemann-Hilbert problem
\begin{eqnarray} \label{NRH2}
\phi^{+}_{j}(p)=\phi^{-}_{j}(p)+R_{j}(\phi^{-}_{1}(p)+\nu_{1}(p),\phi^{-}_{2}(p)+\nu_{2}(p)),
p \in \gamma, j=1,2
\end{eqnarray}
on the unit circle $\gamma$ in the complex $p$- plane, where $\vec{R}(\vec{s})=(R_{1}(s_{1},s_{2}),R_{1}(s_{1},s_{2}))^{T}$ are given differentiable spectral data satisfying the constraint
\begin{eqnarray} \label{constraint2}
\{\mathcal{R}_{1}(s_{1},s_{2}),\mathcal{R}_{2}(s_{1},s_{2})\}_{(s_{1},s_{2})}=1,\nonumber\\
\mathcal{R}_{j}(s_{1},s_{2}) \equiv s_{j}+R_{j}(s_{1},s_{2}),~~~j=1,2
\end{eqnarray}
and $\nu_{j}~(j=1,2)$ are the explicit functions
\begin{eqnarray} \label{nunu}
\vec{\nu}=\left(\begin{array}{c}
            \nu_{1} \\
            \nu_{2}
          \end{array} \right)
          =\left(\begin{array}{c}
            \textbf{i}(\frac{\sqrt{u}}{2} p-S_{z}+\frac{\sqrt{u}}{2} \frac{\delta}{p}) \\
          t(\frac{\sqrt{u}}{2} p-S_{z}-\frac{\sqrt{u}}{2} \frac{\delta}{p})+\frac{z}{2}
          \end{array} \right).
\end{eqnarray}
The vectors $\vec{\phi}^{+}=(\phi^{+}_{1},\phi^{+}_{2})^{T},\vec{\phi}^{-}=(\phi^{-}_{1},\phi^{-}_{2})^{T}$ are unknown vector solutions of the Riemann-Hilbert problem (\ref{NRH2}), analytic respectively inside and outside the contour $\gamma$ such that $\vec{\phi}^{-} \to \vec{0}$ when $p \to \infty$.

If the above nonlinear Riemann-Hilbert problem (\ref{NRH2}) and its linearized form  are uniquely solvable, and if the solutions of (\ref{NRH2}) satisfy the following closure conditions
\begin{subequations} \label{closure2}
\begin{eqnarray}
\sqrt{u}=\lim_{p \to \infty} \delta p\left[\textbf{i}\phi^{-}_{1}(p)+\frac{1}{t}\phi^{-}_{2}(p)\right]
\end{eqnarray}
\begin{eqnarray}
S_{z}=\frac{z-\bar{z}}{4t}-\frac{\textbf{i}}{2} \phi^{+}_{1}(0)+\frac{1}{2t} \phi^{+}_{2}(0),
\end{eqnarray}
\end{subequations}
then $\vec{\psi}^{\pm}=\vec{\phi}^{\pm}+\vec{\nu}$ are common eigenfunctions of the vector fields (\ref{mathcalL}):$\mathcal{L}_{j} \vec{\psi}^{\pm}=\vec{0}~(j=1,2)$ satisfying the relations
\begin{eqnarray}
\{\psi^{\pm}_{1},\psi^{\pm}_{2}\}_{(p,z)}=\frac{\textbf{i}}{4} \sqrt{u}
\end{eqnarray}
and the potentials $u,S_{z}$ reconstructed through (\ref{closure2}) solve the dDS-II system.

In addition, if the spectral data $\vec{\mathcal{R}}({\vec{s}})$ satisfy the reality constraint
\begin{eqnarray} \label{reality}
\vec{\mathcal{R}}\left(\overline{\vec{\mathcal{R}}(\overline{\vec{s}})}\right)=\vec{s},~~\forall \vec{s} \in \mathbb{C}^{2},
\end{eqnarray}
then the eigenfunctions $\vec{\psi}^{\pm}(p)$ satisfy the following symmetry relation
\begin{eqnarray}
\vec{\psi}^{+}(p)=\overline{\vec{\psi}^{-}(-\delta/\bar{p})},
\end{eqnarray}
and the solutions $u,S \in \mathbb{R}$ . Here $\delta=1$ is corresponding to the focusing case and $\delta=-1$ is corresponding to the defocusing case.

\textbf{Proof.} The Riemann-Hilbert problem (\ref{NRH2}) could be formulated directly in terms of the eigenfunctions $\vec{\psi}^{\pm}=\vec{\phi}^{\pm}+\vec{\nu}$ as follows
\begin{eqnarray}  \label{NRHeigen2}
\psi^{+}_{j}(p)&=&\mathcal{R}_{j}\left(\psi^{-}_{1}(p),\psi^{-}_{2}(p)\right) \nonumber\\
&=&\psi^{-}_{j}(p)+R_{j}\left(\psi^{-}_{1}(p),\psi^{-}_{2}(p)\right),\lambda \in \Gamma, j=1,2
\end{eqnarray}
with the normalization
\begin{eqnarray} \label{normalization2}
\vec{\psi}^{-}(p)=\vec{\nu}(p)+O(p^{-1}),~~~|p| >>1.
\end{eqnarray}
By applying the operators  $\mathcal{L}_{j} (j=1,2)$  defined in (\ref{mathcalL}) to the nonlinear Riemann-Hilbert problem (\ref{NRHeigen2}), one obtains the linearized Riemann-Hilbert problem
\begin{eqnarray}
\mathcal{L}_{j}\vec{\psi}^{+}(p)=J \mathcal{L}_{j}\vec{\psi}^{-}(p),~~p \in \gamma.
\end{eqnarray}
where $J$ is the Jacobian matrix defined as $J_{ij}=\partial \mathcal{R}_{i}/\partial s_{j}~(i,j=1,2).$ Based on the normalization (\ref{normalization2}),we have $\mathcal{L}_{j}\vec{\psi}^{-}(p) \to \vec{0}~(j=1,2)$ when $p \to \infty$. It follows that with the uniquely solvable assumption, $\vec{\psi}^{\pm}$ are common eigenfunctions of the vector fields, i.e., $\mathcal{L}_{j} \vec{\psi}^{\pm}=\vec{0}~(j=1,2).$   Consequently, the potentials $u,S$  solve the dDS-II system.
Then the eigenfunctions exhibit the following asymptotic expressions
\begin{subequations} \label{asymptotic2}
\begin{eqnarray}
\psi^{-}_{1}(p)=\textbf{i}\left(\frac{\sqrt{u}}{2} p-S_{z}+\frac{V}{2 \sqrt{u}} \frac{-\delta}{p}\right)+O(p^{-2}),~~~~|p| >>1,
\end{eqnarray}
\begin{eqnarray}
\psi^{+}_{1}(p)=-\textbf{i}\left(\frac{\sqrt{u}}{2} \frac{-\delta}{p}-S_{\bar{z}}+\frac{W}{2 \sqrt{u}} p\right)+O(p^{2}),~~~~|p| <<1,
\end{eqnarray}
\begin{eqnarray}
\psi^{-}_{2}(p)=\frac{t \sqrt{u}}{2} p+(\frac{z}{2}-tS_{z})+\frac{tV}{2 \sqrt{u}} \frac{-\delta}{p}+O(p^{-2}),~~~~|p| >>1,
\end{eqnarray}
\begin{eqnarray}
\psi^{+}_{2}(p)=\frac{t \sqrt{u}}{2} \frac{-\delta}{p}+(\frac{\bar{z}}{2}-tS_{\bar{z}})+\frac{tW}{2 \sqrt{u}} p+O(p^{2}),~~~~|p| <<1,
\end{eqnarray}
\end{subequations}
with $W_{z}=u_{\bar{z}},~V_{\bar{z}}=u_{z}$, implying the closure conditions (\ref{closure2}) which can be viewed as a nonlinear system of two algebraic equations for $u$ and $S_{z}$.

Based on the relation $\{\psi^{+}_{1},\psi^{+}_{2}\}_{(p,z)}=\{\mathcal{R}_{1},\mathcal{R}_{2}\}_{(\psi^{-}_{1},\psi^{-}_{2})} \{\psi^{-}_{1},\psi^{-}_{2}\}_{(p,z)}, p \in \gamma$ from (\ref{NRHeigen2}), the constraint (\ref{constraint2}) implies that
$\{\psi^{+}_{1},\psi^{+}_{2}\}_{(p,z)}=\{\psi^{-}_{1},\psi^{-}_{2}\}_{(p,z)},p \in \gamma$, i.e. the Poisson brackets $\{\psi^{+}_{1},\psi^{+}_{2}\}_{(p,z)},\{\psi^{-}_{1},\psi^{-}_{2}\}_{(p,z)}$ are analytic in the whole complex $p$-plane. As $\{\psi^{-}_{1},\psi^{-}_{2}\}_{(p,z)} \to \frac{\textbf{i}}{4} \sqrt{u}$ when $p \to \infty$, it follows that $\{\psi^{+}_{1},\psi^{+}_{2}\}_{(p,z)}=\{\psi^{-}_{1},\psi^{-}_{2}\}_{(p,z)}=\frac{\textbf{i}}{4} \sqrt{u}$.

In addition, by taking the vector $\vec{s}=\overline{\vec{\psi}^{-}(-\delta/\bar{p})}$ (with $\delta=\pm 1$) into the reality constraint (\ref{reality}), one obtains the following relation directly from the uniquely solvable assumption
\begin{eqnarray} \label{unitcircle}
\vec{\psi}^{+}(p)=\overline{\vec{\psi}^{-}(-\delta/\bar{p})},~~~|p| \leq 1.
\end{eqnarray}
By taking this relation (\ref{unitcircle}) into the asymptotic expressions (\ref{asymptotic2}), we have the  reality condition $u,S \in \mathbb{R}$.

~~~~~~~~~~~~~~~~~~~~~~~~~~~~~~~~~~~~~~~~~~~~~~~~~~~~~~~~~~~~~~~~~~~~~~~~~~~~~~~~~~~~~~~~~~~~~~~~~~~$\square$

\textbf{Remark 6.} By introducing the parameter
\begin{eqnarray}
p=e^{-\textbf{i} \theta},~~~\theta \in \mathbb{R}
\end{eqnarray}
the nonlinear Riemann-Hilbert problem (\ref{NRH2}) can be characterized by the following system of nonlinear integral equations
\begin{eqnarray}
\phi^{\pm}_{j}(e^{-\textbf{i}\theta})&=&\frac{1}{2\pi} \int^{2\pi}_{0} \frac{~d\theta'}{1-(1\mp \varepsilon)e^{-\textbf{i}(\theta'-\theta)}}R_{j}[\phi^{-}_{1}(e^{-\textbf{i}\theta'})+\nu_{1}(e^{-\textbf{i}\theta'}),\phi^{-}_{2}(e^{-\textbf{i}\theta'})+\nu_{2}(e^{-\textbf{i}\theta'})] \nonumber\\
 &&
\end{eqnarray}
and the closure conditions (\ref{closure2}) read as
\begin{subequations} \label{closureintegral}
\begin{eqnarray}
\sqrt{u}&=&\lim_{p \to \infty} \delta p\left[\textbf{i}\phi^{-}_{1}(p)+\frac{1}{t}\phi^{-}_{2}(p)\right] \nonumber\\
&=&-\frac{\delta}{2\pi} \int^{2\pi}_{0}\{\textbf{i}R_{1}[\phi^{-}_{1}(e^{-\textbf{i}\theta})+\nu_{1}(e^{-\textbf{i}\theta}),\phi^{-}_{2}(e^{-\textbf{i}\theta})+\nu_{2}(e^{-\textbf{i}\theta})] \nonumber\\
&&+\frac{1}{t}~R_{2}[\phi^{-}_{1}(e^{-\textbf{i}\theta})+\nu_{1}(e^{-\textbf{i}\theta}),\phi^{-}_{2}(e^{-\textbf{i}\theta})+\nu_{2}(e^{-\textbf{i}\theta})]\}~e^{-\textbf{i}\theta} d\theta, \nonumber\\
&&
\end{eqnarray}
\begin{eqnarray}
S_{z}&=&\frac{z-\bar{z}}{4t}-\frac{\textbf{i}}{2} \phi^{+}_{1}(0)+\frac{1}{2t} \phi^{+}_{2}(0) \nonumber\\
&=&\frac{z-\bar{z}}{4t}+\frac{1}{4\pi} \int^{2\pi}_{0}\{-\textbf{i}R_{1}[\phi^{-}_{1}(e^{-\textbf{i}\theta})+\nu_{1}(e^{-\textbf{i}\theta}),\phi^{-}_{2}(e^{-\textbf{i}\theta})+\nu_{2}(e^{-\textbf{i}\theta})] \nonumber\\
&&+\frac{1}{t}~R_{2}[\phi^{-}_{1}(e^{-\textbf{i}\theta})+\nu_{1}(e^{-\textbf{i}\theta}),\phi^{-}_{2}(e^{-\textbf{i}\theta})+\nu_{2}(e^{-\textbf{i}\theta})]\} d\theta, \nonumber\\
&&
\end{eqnarray}
\end{subequations}
in which the explicit functions $\nu_{1}(p),\nu_{2}(p)$ are defined in (\ref{nunu}).

~~~~~~~~~~~~~~~~~~~~~~~~~~~~~~~~~~~~~~~~~~~~~~~~~~~~~~~~~~~~~~~~~~~~~~~~~~~~~~~~~~~~~~~~~~~~~~~~~~~$\square$

\section{Outlook}

These results are the necessary background for all the future
studies we are planning to make, and that consist of the following steps.

1) The use the Manakov-Santini method for vector fields to study the Cauchy problem for the dDS (dispersionless Davey-Stewartson) system (\ref{dDS}), based on the above results, through the following: the identification of the appropriate formal zero energy eigenfunctions of the vector fields Lax pair, analytic
in the complex parameter respectively in a neighborhood of $\infty$ and $0$; the identification of the vector nonlinear Riemann-Hilbert inverse problem on a closed curve in the complex plane.

2) The use of the nonlinear Riemann-Hilbert inverse problem to construct the longtime behavior of the solutions of the Cauchy problem.

3) The use of the Riemann-Hilbert inverse problem to construct classes of exact implicit solutions of the dDS system  (\ref{dDS}).

4) The use of the  Riemann-Hilbert inverse problem to study how the above dynamics give rise to a wave breaking, and to
study analytically the nature of such wave breaking, with the identification of the type of
singularities, in the focusing as well as in the defocusing cases.

\noindent
\bigskip
\bigskip
\bigskip

\textbf{Acknowledgements:} This work has been supported by the National Natural Science Foundation of China (No. 11501222). The author would like to thank Professor Paolo Maria Santini very sincerely for the valuable suggestions and patient guidance to accomplishment of this work.

\end{document}